\documentclass{pasj00}

\begin{document}
\SetRunningHead{H.~Imai et~al.~}{H$_2$O masers in IRAS 22480$+$6002}
\Received{2007/05/30}
\Accepted{2007/09/18 ; Ver. 3.0 Sept. 18, 2007}
\SetVolumeData{2008}{60}{1}
\Published{2008, No. 1, Feb. 25 issue in press}
\title{JVN observations of H$_2$O masers around the evolved star IRAS~22480$+$6002}
\author{Hiroshi \textsc{Imai}\altaffilmark{1},
Takahiro \textsc{Fujii}\altaffilmark{1,2}, 
Toshihiro \textsc{Omodaka}\altaffilmark{1}}
\and
\author{Shuji \textsc{Deguchi}\altaffilmark{3}}
\altaffiltext{1}{Department of Physics, Faculty of Science, \\
Kagoshima University, 1-21-35 Korimoto, Kagoshima 890-0065}
\altaffiltext{2}{VERA Project Office, National Astronomical Observatory, \\
2-21-1 Osawa, Mitaka, Tokyo 181--8588}
\altaffiltext{3}{Nobeyama Radio Observatory, 
National Astronomical Observatory,\\
Minamimaki, Minamisaku, Nagono 384-1305} 
\email{(HI: hiroimai@sci.kagoshima-u.ac.jp)}
%\Published{2007/??/??}
%

%\setlength{\baselineskip}{6ex}

\maketitle

\begin{abstract}
We report on the H$_2$O maser distributions around IRAS 22480+6002 (=IRC+60370) 
observed with the Japanese VLBI Network (JVN) at three epochs spanning 2 months. 
This object was identified as a K-type supergiant in 1970s,  
which was unusual as a stellar maser source. The spectrum of H$_2$O masers
consists of 5 peaks separated roughly equally by a few km s$^{-1}$ each. The H$_2$O masers 
were spatially resolved into more than 15 features, which spread about 50 mas along the east--west direction. 
However, no correlation was found between the proper motion vectors and their spatial distributions; 
the velocity field of the envelope seems random. 
A statistical parallax method applied to the observed proper-motion data set 
gives a distance of $1.0\pm 0.4$ kpc for this object, that is considerably smaller than previously thought. 
The distance indicates that this is an evolved star with $L\sim 5800\ L_{\odot}$.
This star shows radio, infrared, and optical characteristics quite similar 
to those of the population II post-AGB stars such as RV Tau variables.  
\end{abstract} 

\section{Introduction}
\label{sec:introduction}
H$_2$O maser emission has been observed in circumstellar envelopes of evolved stars such as O-rich Mira variables 
and OH/IR stars with large mass loss rates of $\dot{M} \geq 10^{-7}M_{\odot}$~yr$^{-1}$ \citep{rei81,eli92}. 
Most of these stars are asymptotic giant branch (AGB) stars or red supergiants both with the spectral type M, 
with a few exceptions for transient stars at pre-planetary nebula phase 
(or supposedly a few pre-main sequence stars such as Ori KL; \cite{mor98}). 
For the central stars with spectral types earlier than M, UV radiation 
from stellar chromosphere eventually dissociates most of molecules (except CO) 
in the inner envelope (e.g., \cite{wir98}). 
Therefore, H$_2$O (or SiO) masers are usually not expected for these stars, 
except for the case that the molecules in dense circumstellar clumps shield themselves from UV radiation. 
In fact, H$_2$O masers found 
in a young planetary nebula \citep{mir01,sue07} must be such an exceptional case. 
OH masers have been found in yellow hypergiants with spectral types F and G (such as IRC+10420 and V1427 Aql) 
\citep{gig76,ned92,hum02}. 
However, H$_2$O  and SiO masers have never been detected in these objects \citep{nak03}, though
thermal emission of a few other molecules have been observed in the outer circumstellar shell \citep{cas01,tey06}.   

%%%%%%%%%%% Put here Tables 1 %%%%%%%%%%%%%
%\setlength{\baselineskip}{6ex}
% \clearpage 
%%%%%%%%%%% Table 1 %%%%%%%%%%%%%%%%%%%
\begin{table*}[ht]
\caption{Status of the telescopes, data reduction, and resulting performances in 
the individual epochs of the JVN observations.}\label{tab:status}
\begin{center}
%%%% Only for submitted version
% \scriptsize
\footnotesize
\begin{tabular}{lccccccc} \hline \hline
& Epoch in & & & Reference & 1-$\sigma$ level & Synthesized & Number of \\
Observation & the year & Duration & Used    
& velocity\footnotemark[2] & noise & beam\footnotemark[3] & detected \\
code & 2005 & (hr) & telescopes\footnotemark[1]  
& (km s$^{-1}$) & (Jy beam$^{-1}$) & (mas)
& features \\ \hline
r05084b \dotfill & March 25 & 7.3 & MZ, IR, OG, IS, KS, NB\footnotemark[4] 
& $-52.3$  & 0.22 & 1.7$\times$1.6, $-$37$^{\circ}$ & 20 \\
r05116b \dotfill & April 26 & 7.3 & MZ, IR, OG, IS\footnotemark[5], KS, NB 
& $-52.0$  & 0.15 & 3.8$\times$2.0, $-$14$^{\circ}$ & 17 \\
r05151a \dotfill & May 31 & 8.1 & MZ, OG\footnotemark[5], IS\footnotemark[5], KS, NB 
& $-$52.6 & 0.15 & 3.2$\times$2.8, $-$66$^{\circ}$ & 14 \\ \hline
\end{tabular}
\end{center}
\footnotemark[1] Telescopes that were effectively operated and whose recorded data were valid: 
MZ: the VERA 20~m telescope at Mizusawa, IR: the VERA 20~m telescope at Iriki, OG: the VERA 20~m telescope at Ogasawara Is., IS: the VERA 20~m telescope at Ishigakijima Is., KS: the NiCT 34-m telescope at Kashima, NB: the NRO 45-m telescope at Nobeyama. \\
\footnotemark[2] Velocity channel used for the phase reference in data 
reduction. \\
\footnotemark[3] The synthesized beam made in natural weight; 
major and minor axis lengths and position angle. \\
\footnotemark[4] Ceasing operation for 2.5~hr due to strong winds and pointing correction. \\
\footnotemark[5] High system temperature ($>$300~K) due to bad weather conditions.
\end{table*}
\ \\

The optical counterpart of IRAS 22480+6002 ($=$AFGL~2968, or IRC$+$60370) 
was identified as a K-type supergiant (K0Ia; \cite{hum74}, or
K4.5Ia; \cite{faw77}). Therefore, the detections of H$_2$O and SiO masers  
\citep{han98,nym98} were surprising. Though a search for OH 1612 MHz emission was negative \citep{les92}, 
CO emission was detected toward this star \citep{jos98}. 
% It indicates that the stellar wind has already blown more than $10^3$ years before. 
From the CO $J=2$--1 line profile, the systemic stellar velocity and 
the expansion velocity of this star were obtained to be 
$V_{\rm lsr}$=$-49.3$ km s$^{-1}$ and $V_{exp}=26.4$ km~s$^{-1}$, respectively 
\citep{jos98, gro99}. They are consistent with those obtained from the H$_2$O\ and SiO maser spectra, 
and the expansion velocity of the envelope of this star is typical for OH/IR stars. 
The radial velocity gives a kinematic distance of 5.0~kpc. It suggests a large luminosity 
$L_{\ast}=$140 000~$L_{\odot}$ of the central star \citep{gro99}, 
but it is consistent with the supergiant interpretation of this object. 
A blue nearby star, a B5II star, is seen by about 12$''$ east of this object.
Though it is cataloged as a visual binary \citep{wor97},
a physical association of this object with the maser source is questionable  
because of the large velocity difference of about 40 km s$^{-1}$ \citep{hum74}.
\citet{win94} gave a new spectral classification of M0I for IRAS 22480+6002
from the low-resolution spectrum between 6000 and 8800 A, which was significantly different from
the previous type assignment of this star. For a long-period variable,
optical spectral classification may vary with light variations.
However, this star has not been reported as a variable star, though it is optically not very faint
($V\sim 8.3$).

In this work, we report three-epoch VLBI observations of H$_2$O\ masers of IRAS 22480+6002 
to rectify the entangled situation associated with this object. 
From the spatio-kinematics of the masers, we diagnose a probable anomaly of a hot wind from the K-type star. 
We estimated the distance to this star using the statistical parallax method 
based on the proper motion data of H$_2$O masers. Our result gives a much smaller distance
for this star than previously thought. The new estimation of the distance demands 
to reconsider various properties of this star. 
Based on the arguments presented in section 3, 
we conclude that this star is a population II post-AGB star.

\section{Observations and Data Reduction}
\label{sec:observation}

The 22 GHz VLBI observations were made at three epochs during 2005 March--May, 
using six telescopes of the Japanese VLBI Network\footnote
{JVN involves the NRO 45-m telescope at Nobeyama, the NiCT 34-m telescope at Kashima, and 4 VERA 20-m telescopes.   
NRO and VERA observatories are branches of the National Astronomical Observatory of Japan, 
a member institute of the inter-university research agency, NINS.}. 
Table \ref{tab:status} gives a summary of status of the observations. 
At each epoch, the observation was made for $\sim$7.5 hours including scans 
at the object and the calibrator (J2202$+$4216). 
The signal was recorded with a rate of 128 Mbit s$^{-1}$ and in two baseband channels 
with a band width of 16 MHz each. The VERA telescopes also simultaneously observed 
the position reference source J2254$+$6209 
with the object using the dual beam system (e.g., \cite{hon03}), 
but the reference source was not detected. 

The data reduction, and identification of maser spots and features  
were made using the NRAO AIPS package in the same setup and procedures as those described in \citet{ima06} and \citet{ino07}. 
The spectral channel spacing was set to 0.21 km~s$^{-1}$. A typical size of the synthesized beam was 2.5~milliarcseconds (mas) 
in the three observations (see Table \ref{tab:status}). A relative position accuracy of a maser spot 
(a component which appears in a single velocity channel)
was typically 0.05~mas depending on a signal-to-noise ratio and spatial structure of the spot. 
A relative position accuracy of a maser feature (a cluster of maser spots which compose a physical clump)
was typically 0.15 mas. Relative proper motions were measured for maser features, 
which were identified at least at two of the three epochs. 
Table 2 gives a list of relative positions of these maser features, 
their proper motions, radial velocities, and peak intensities.

Using the fringe-rate mapping method, we estimated the absolute coordinates of the maser feature IRAS 22480$+$6002: I2007 {\it 2} 
(the $V_{\rm lsr}=-58.48$ km s$^{-1}$ component) to be  
R.A.(J2000.0)$=$22$^{h}$49$^{m}$58$^{s}$\hspace{-2pt}.876, and
decl.(J2000.0)$=+$60\arcdeg17\arcmin56\arcsec.65, with uncertainty of $\sim$0\arcsec.1. 
The feature location is coincident with the 2MASS position of the bright star J22495897+6017568, 
or with the GSC 2.2 position of  N012302336407,
%(22$^{h}$49$^{m}$58$^{s}$\hspace{-2pt}.97, 60\arcdeg17\arcmin56\arcsec.8) 
within 0.8$''$ and 0.6$''$, respectively.

%%%%%%%%%%%%%%%%%%%%%%%%%%%%%%%%%%%%%%%%% section 3 %%%%%%%%%%%%%%%%%%%%%%%%%%%%%%

%%%%%%%% Figure 1 %%%%%%%%%%%%%%%%%
\begin{figure}[htb]
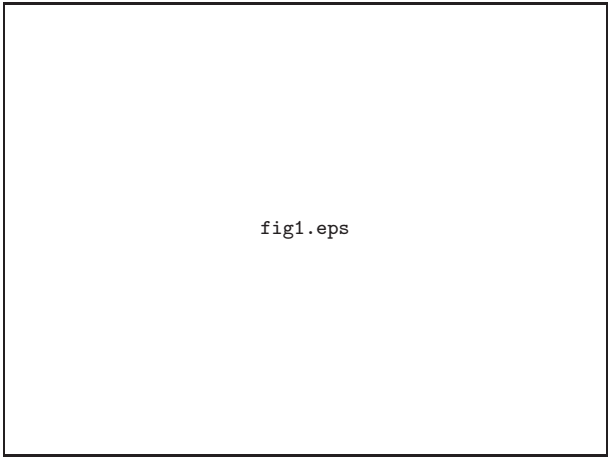

  \begin{center}
     \FigureFile(8cm,6cm){fig1.eps}
  \end{center}
  \caption{Cross-power spectra of H$_2$O\ masers in IRAS 22480$+$6002 obtained in the baseline 
between the NRO~45m and NICT~34m telescopes at three epochs.}
\label{fig:I2248_spectrum}
\end{figure}
%%%%%%%

\section{Results and discussion}

%%%%%%%%%%% Put here Tables 2 %%%%%%%%%%%%%
%%%%%%%%%%% Put here figures 2,3 %%%%%%%%%%%%%

%%%%%%%%%%% Table 2 %%%%%%%%%%%%%%%%%%%
\begin{table*}[ht]
\caption{Parameters of the H$_2$O  maser features identified by 
proper motion toward IRAS 22480$+$6002.} \label{tab:pmotions}
\begin{center}
%{\scriptsize
%{\footnotesize
\begin{tabular}{lrrrrrrrrrrr} \hline \hline    
Feature\footnotemark[1]
 & \multicolumn{2}{c}{Offset}
 & \multicolumn{4}{c}{Proper motion\footnotemark[2]}
 & \multicolumn{2}{c}{Radial motion\footnotemark[3]}
 & \multicolumn{ 3}{c}{Peak intensity at  3 epochs} \\                           
 & \multicolumn{2}{c}{(mas)} 
 & \multicolumn{4}{c}{(mas yr$^{-1}$)}
 & \multicolumn{2}{c}{(km s$^{-1}$)}
 & \multicolumn{ 3}{c}{(Jy beam$^{-1}$)} \\                                    
 & \multicolumn{2}{c}{\hrulefill} 
 & \multicolumn{4}{c}{\hrulefill} 
 & \multicolumn{2}{c}{\hrulefill} 
 & \multicolumn{ 3}{c}{\hrulefill} \\                                       
  & $\Delta$R.A. & $\Delta$decl. & $\mu_{x}$ & $\sigma(\mu_{x})$ & $\mu_{y}$ 
 & $\sigma(\mu_{y})$ & $V_{z}$ & $\Delta V_{z}$\footnotemark[4]
 & Epoch  1& Epoch  2& Epoch  3 \\ \hline                                        
  1   \ \dotfill \  &$    -6.56$&$   -10.69$&$   4.20$&   1.46 &$   5.86$&   1.40
 &$ -59.78$&   0.26  &        0.19 &        0.17 &    ...        \\                                 
  2   \ \dotfill \  &$     0.00$&$     0.00$&$   0.00$&   0.36 &$   0.00$&   1.29
 &$ -58.48$&   2.74  &        2.33 &        2.72 &        3.48   \\                                 
  3   \ \dotfill \  &$   -45.23$&$     7.26$&$  -0.87$&   0.42 &$   0.51$&   0.66
 &$ -55.01$&   2.04  &        3.77 &        2.47 &        2.42   \\                                 
  4   \ \dotfill \  &$    -4.95$&$     7.17$&$   0.25$&   0.29 &$   0.47$&   1.44
 &$ -54.33$&   1.05  &        0.41 &        0.53 &        0.73   \\                                 
  5   \ \dotfill \  &$   -26.70$&$    -2.86$&$  -1.03$&   0.97 &$  -0.72$&   0.61
 &$ -52.34$&   3.37  &        8.11 &        8.73 &        8.16   \\                                 
  6   \ \dotfill \  &$    -0.37$&$     1.50$&$   3.96$&   1.29 &$   0.86$&   1.91
 &$ -50.37$&   0.56  &        0.53 &        0.73 &        0.49   \\                                 
  7   \ \dotfill \  &$     0.16$&$     2.74$&$   0.05$&   0.61 &$   0.98$&   1.56
 &$ -49.29$&   2.18  &        8.44 &        7.96 &        6.20   \\                                 
  8   \ \dotfill \  &$    -0.04$&$    13.96$&$   0.90$&   0.67 &$  -0.29$&   0.78
 &$ -47.88$&   1.33  &        1.92 &        1.69 &        1.03   \\                                 
  9   \ \dotfill \  &$   -34.92$&$    -7.63$&$  -1.95$&   1.66 &$  -3.07$&   2.97
 &$ -47.39$&   0.52  &        0.25 &        0.26 &    ...        \\                                 
 10   \ \dotfill \  &$   -48.23$&$    11.08$&$   2.27$&   1.18 &$   0.71$&   0.88
 &$ -47.24$&   0.56  &        0.24 &        0.20 &        0.18   \\                                 
 11   \ \dotfill \  &$    -2.24$&$     8.18$&$   0.06$&   1.10 &$  -0.16$&   0.64
 &$ -46.17$&   0.84  &        0.41 &        0.28 &        0.26   \\                                 
 12   \ \dotfill \  &$    -1.02$&$    13.29$&$   0.06$&   0.41 &$   0.56$&   0.74
 &$ -45.37$&   1.97  &        3.34 &        3.40 &        2.44   \\                                 
 13   \ \dotfill \  &$     0.13$&$     7.77$&$  -1.86$&   3.19 &$   0.26$&   1.32
 &$ -44.40$&   0.77  &        0.24 &        0.33 &        0.32   \\                                 
 \hline
 \end{tabular}
% }
 \end{center}
\noindent
\footnotemark[1]
H$_2$O maser features detected toward IRAS 22480+6002. The feature is designated as 
IRAS 22480+6002:I2007 {\it N}, where {\it N} is the ordinal source number given in this 
column (I2007 stands for sources found by Imai et~al. and listed in 2007). \\
\footnotemark[2]
Relative value with respect to the motion of the position-reference maser feature: 
IRAS 22480+6002:I2007 {\it 2}. \\
\footnotemark[3] Relative value with respect to the local stand of rest. \\
\footnotemark[4] Mean full velocity width of a maser feature at half intensity. 
\end{table*}
%%%%%%%%%

\subsection{Spatial distribution and proper motions of maser features}

\begin{figure}[htb]
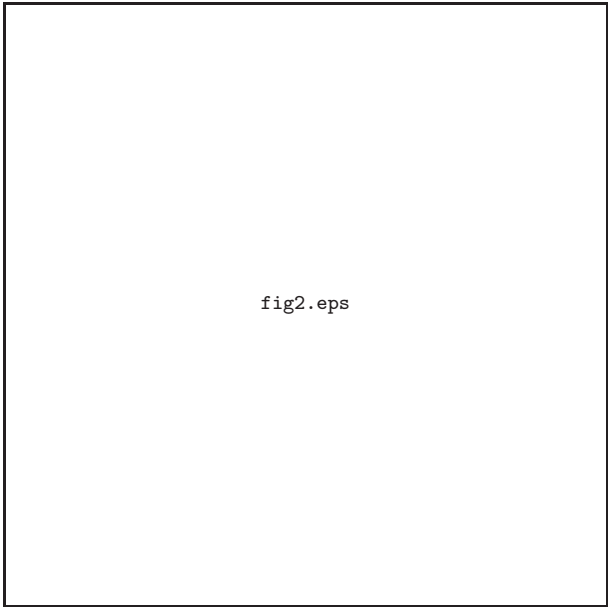

  \begin{center}
     \FigureFile(8cm,8cm){fig2.eps}
  \end{center}
  \caption{Distribution of H$_2$O masers on 2005 March 25. The color code indicates the radial velocity of the feature, 
  and the size of the filled circle indicates the flux density of the feature. Note that the $-58.48$ km s$^{-1}$ reference component 
  is located at the origin (light blue), but is almost overlapped with the systemic-velocity component 
  ($\sim -49$ km s$^{-1}$) shown in green.}
\label{fig:1st-epoch}
\end{figure}

Figure \ref{fig:I2248_spectrum} shows cross-power spectra of the H$_2$O masers of IRAS 22480+6002. 
The H$_2$O maser emission spread in a velocity range of 15 km~s$^{-1}$, which is typical 
for Mira-type AGB stars (e.g., \cite{tak94}). Five spectral peaks were seen in roughly equal separations of 2--3 km~s$^{-1}$; 
the second highest peak was near the systemic velocity ($V_{\rm lsr}=-49.3$ km s$^{-1}$). 
The correlated powers of these peaks equally increased by a factor of two 
during 2 months in our observing run, except for the second peak 
for which the intensity increased only by about 20\%. 
However, the peak flux densities of individual features 
were found not to vary much (Table 2). This fact indicates that extended emissions
were partially resolved in the shortest baseline between NRO and NICT (197.4 km),
but resolved-out in the longer baselines. % baseline length 197385.078 m
Correlated flux densities were estimated to be
about 30--40\% of the total-power intensities. 
Figure \ref{fig:1st-epoch} shows the distribution
of maser features at the first epoch.  
The extent of 50 mas corresponds to 50~AU ($D$/1 kpc), which is somewhat larger than 
those seen around Mira variables at $D=1$ kpc (e.g., \cite{bow94}). 
In this figure, one of the low-velocity (blue-shifted) components
($V_{\rm lsr}=-58.48$ km s$^{-1}$: the position reference) is located at the origin 
and the other low velocity components 
(in grey and blue colors) are located both near the eastern and western edges. 
Many of the higher velocity (red-shifted)  components
(shown in yellow, orange, and red colors) fall at the eastern edge, 
but a few of them are scattered at the west side too.  
The overall distribution of water maser features are characterized by the elongation to the
east-west direction. But, no clear correlation is found between the velocities and spatial positions.   
If the circumstellar envelope of the K supergiant interacts with the wind from the eastern BII star
(though this is unlikely), maser spots and features could be aligned perpendicularly to the wind direction, 
i.e., in the north--south direction (e.g., \cite{ima02b}). We find no such N--S alignment of the H$_2$O maser features. 
\citet{mei99} noted that the MIR image of this star with the NASA 3-m telescope showed
a northeast-southwest elongation, but concluded that it was likely to be an artifact caused by astigmatism.  

%%%%%%%% Figure 2 %%%%%%%%%%%%%%%%%

\begin{figure}[ht]
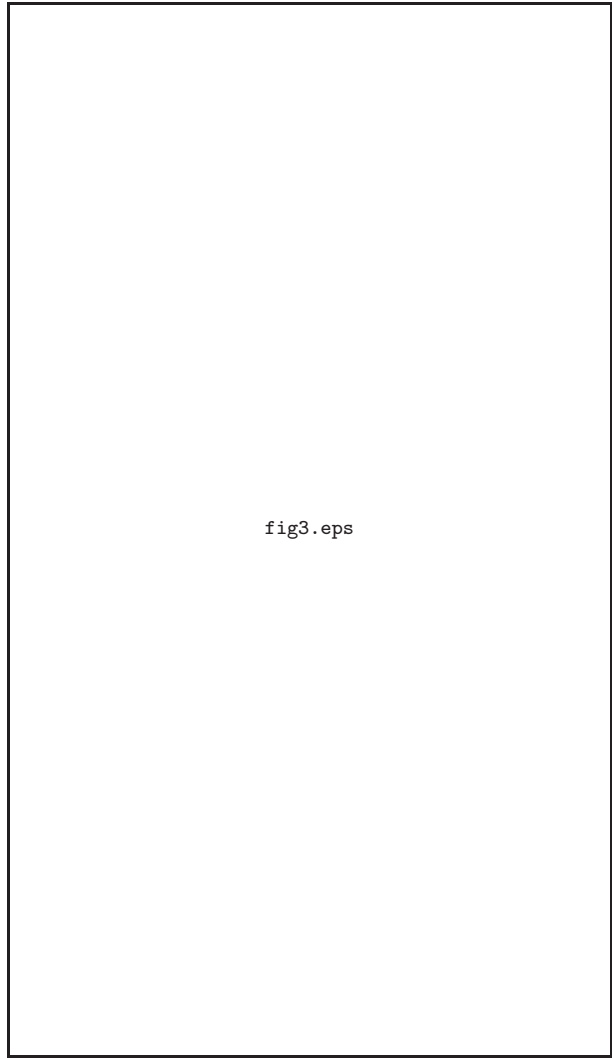

%  \begin{center}
    \FigureFile(8cm,14cm){fig3.eps}
%  \end{center}
\caption{Observed relative proper motions of H$_2$O  maser features in IRAS 22480$+$6002 in R.A. (a) 
and decl. (b) directions. 
The number on the left indicates the feature name designated in Table 2. 
The vertical bar attached to each data point indicates the position uncertainty.
The least-square-fitted line is also shown.}
\label{fig:PM-I2248}
\end{figure}

We detected 14--20 H$_2$O maser features though all epochs  (the last column of table \ref{tab:status}).  
Note that the H$_2$O masers were persistent in velocity and in spatial distribution during the two months; 
65--90\% of the detected maser features survived during our observing run. 
Therefore, we identified the same maser features at three epochs relatively easily, 
and measured the proper motions of the individual features during two months.
Table \ref{tab:pmotions} gives the measured proper motions. 
Figure \ref{fig:PM-I2248} shows the linear fits to the relative positions 
of the individual maser features.
%%%%%%%%% inserted further %%%%%%%%%
% {\bf 
The fitted proper motions look significant with the second
epoch contributing little for most features, which warrants our selecting 
the same maser features at the different epochs. %} 
%%%%%%%%%  inserted %%%%%%%%%
Circumstellar H$_2$O masers can amplify the radiation of the central star 
(for example, see the case of U Her; \cite{vle02}). % though Bains et al. 2003 does not conrim this.
In the present case, we may speculate that the $-52.34$ km s$^{-1}$ feature
(No. 5 in Table 2 and Figure 4), which is one of the strongest components 
and located near the center of the maser distribution, is such a maser amplifying the steller radiation. 
However, it is hasty to draw any conclusion from this observation, 
since we have no information on the central-star position in this scale.

%%%%%%%% Figure 3 %%%%%%%%%%%%%%%%%

%%%%%%%% Figure 4 %%%%%%%%%%%%%%%%%
\begin{figure}[ht]
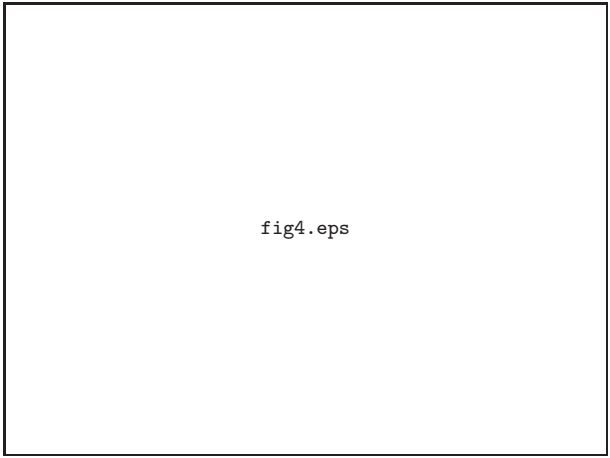

%  \begin{center}
    \FigureFile(8cm,6cm){fig4.eps}
%  \end{center}
  \caption{Doppler velocities (colorfully displayed) and relative proper motion vectors 
  (indicated by arrows) of H$_2$O  masers in IRAS 22480$+$6002. The displayed proper motion vector is that 
  subtracted by a velocity bias ($\overline{\mu _{x}}, 
\overline{\mu _{y}})=(0.97, 0.72)$ [mas yr$^{-1}$] from the original vector 
to cancel out the average motions of all the features.
A number added for each feature with a proper motion shows the assigned one 
after the designated name form ``IRAS 22480$+$6002: I2007". The map origin is set to the location 
of the feature IRAS 22480$+$6002: I2007 {\it 2}.}
\label{fig:I2248-velocity}
\end{figure}
%\clearpage

Figure \ref{fig:I2248-velocity} shows the proper motion vectors of the individual H$_2$O maser features.
%In this figure, the motion of the reference feature, (0.97 mas yr$^{-1}$, 0.72 mas yr$^{-1}$), 
% was introduced to cancel out the average motions of all the features.
Note that the largest two proper-motion vectors at the lower left and lower right,  
i.e., features 1 and 9 of the $V_{\rm lsr}=-59.78$ and $-47.39$ km s$^{-1}$ components, respectively, 
were determined by two-epoch detections, 
so that they are slightly inaccurate. The proper motions of all other features with 3-epoch detections are 
within a few mas per year (relative to the reference component at $V_{\rm lsr}=-58.48$ km s$^{-1}$). 
%Therefore, the distance to this object is expected roughly 1--2 kpc, if the magnitude of transverse motions 
%is similar to the expansion motion of about 25 km s$^{-1}$. However, 
We cannot find any systematic trend of motions in this diagram.
%\footnote{
For example, features 3 and 10 at the western edge move in opposite directions, 
and features 6 and 13 at the eastern edge also move in opposite directions.

\begin{figure}[ht]
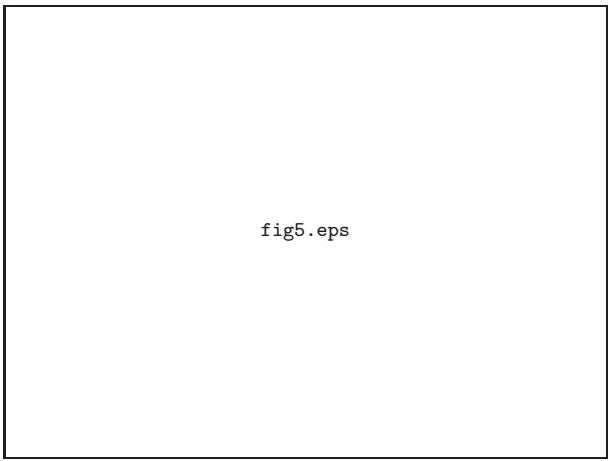

%  \begin{center}
    \FigureFile(8cm,6cm){fig5.eps}
%  \end{center}
  \caption{Plot of relative R.A. offset (left) and proper motion (right) against radial velocity. 
The filled and unfilled circles indicate
the three-epoch and two-epoch detections. The large and small ellipses in both panels 
indicate the position and proper motion curves expected from thin spherical shell models 
(one-dimensional in the R.A. direction) with a constant expansion velocity 
of 12.5 km s$^{-1}$ and a radius of $3.5\times 10^{14}$ cm, 
and with 7 km s$^{-1}$ and $8\times 10^{13}$ cm, respectively both at a distance of 0.9 kpc.
The observed points in the right panel do not fit to these ellipses.
}
\label{fig:expansion}
\end{figure}

Figure \ref{fig:expansion} shows the RA-offsets and $\mu_x$ plots against $V_{\rm lsr}$. 
The ellipse is a plot of
expected offset and proper motion from a thin spherical-shell model with a constant velocity 
(in the Right Ascension direction because the maser features are spread mainly in this direction). 
If the shell model is correct, all of the maser features should fall between these ellipses. 
However, the right panel does not show such a tendency. 
Rather, the observed points seem to distribute randomly.

%%%%%%%%%%%% from the footnote %%%%%%%%%%
%{\bf 
The randomness of the proper motions may partially originate from the large
random errors in the position measurements. %, because the positional uncertainties 
% are considerably large (figure 3). 
In order to check this issue,
we made a Monte Carlo simulation of the 3-epoch proper motion fitting 
with the same positional uncertainties but without real motions 
(i.e., position jitters only due to the measurement errors). We obtained
the mean velocity dispersions ($0.79\pm 0.25$ mas yr$^{-1}$, $0.80\pm 0.20$ mas yr$^{-1}$) 
for the 13 proper motions in R.A. and Dec. directions for the present case from the simulations;
here, the number after the '$\pm$' sign is a standard deviation of dispersions obtained in the simulations.
The observed velocity dispersions (1.95 mas yr$^{-1}$, 1.93 mas yr$^{-1}$), 
are significantly larger than the simulated mean dispersions (more than $4 \sigma$). 
Therefore, the observed proper motions are substantially real motions of masing features.
%}

%\clearpage
%%%%%%%% Figure 5 %%%%%%%%%%%%%%%%%

The observed random motions may originate from the intrinsic random ballistic motions 
of matters ejected from or infalling into the atmosphere of the central supergiant. 
%or the interaction of hot wind as mentioned above, though it is difficult to obtain a clear conclusion here. 
This may be a characteristic of mass outflow of a supergiant 
originating from the extended atmosphere which is considerably turbulent \citep{lev05,jos07}.
The motion of 1 mas yr$^{-1}$ corresponds a transverse motion of $\sim 4.7 (D/$kpc) km s$^{-1}$. 
In order to obtain the distance, we applied the statistical parallax method 
to the obtained maser proper motions; %{\bf 
for example, see \citet{sch81}. %}
Assuming random motions of maser features, 
we obtained a velocity dispersion in radial motion 
(with respect to the average velocity of maser features, $V_{\rm lsr}=-50.6$ km s$^{-1}$) 
to be $\sigma_v \simeq 5.0$ km s$^{-1}$, which is smaller 
than the outflow velocity estimated from CO emission.

The dispersion in the maser proper motions can be obtained 
by subtracting the dispersion involved in the measurements; %{\bf 
see equation (3) of \citet{sch81}. %}
We obtain $\sigma _{\mu} \sim 1.40$ mas yr$^{-1}$ and get
a distance to IRAS 22480+6002 
%  floowing change for bold
%{\bf 
to be $D=\sigma_{v}/\sigma_{\mu}= 0.76 \ (\pm 0.25)$ kpc. 
The formal uncertainty involved in the distance estimation was computed using equation (4) of \citet{sch81}.  
If we exclude the largest two proper-motion features with two-epoch detections from the sample,
we get the distance $1.02 \ (\pm 0.38) $ kpc. 
%}
% till here
Later on, we adopt this distance for IRAS 22480+6000,
because large motions detected by two-epoch observations are somewhat dubious.
Note that this distance is derived based on the assumption that the velocity field of
masers is random and isotropic, and that the proper motions appeared in maser features are
real motions of gas clumps. The distance 1.0 kpc gives a radius of water maser shell approximately 
$3.7\times 10^{14}$ cm, which is compatible with the radii of water maser shells of miras,
but considerably smaller than those of M-supergiants \citep{yat94,cot04}. 
Though the obtained distance still involve a considerable uncertainty, 
it excludes the possibility of a very large distance of 5 kpc (a kinematic distance).

%Adopting this distance, the separation of two stars is estimated to be $\sim$5800~AU. 
The luminosity of this star is re-evaluated to be $5.8 \times 10^3~L_{\odot}$ (reestimated from \cite{gro99}).  
It is considerably small for a supergiant.
However, from the distance of 1.0 kpc, we can compute the absolute V magnitude of this star  
from 2MASS K magnitude ($K=2.8$) using $V-K=3.7$ (for K5III; \cite{zom90}), and with reddening corrections,  
we get $M_V\sim -4.4$. This value falls near the absolute magnitude of K5Ib (or M0Ib) \citep{zom90}. 
Therefore, the luminosity is still in a range of supergiants.

The radial velocity of $\sim -50$ km s$^{-1}$ is typical for young objects
in the Perseus spiral arm  in the direction of this star (for example, see \cite{sit03}). 
If we take into account the large uncertainties involved in the obtained distance, we cannot completely exclude
the possibility that IRAS 22480+6002 belongs to the Perseus arm 
at $D\sim 3.0$ kpc at $l=108^{\circ}$ \citep{xux06}.
However,  there are  several other bright stellar maser sources with similar radial velocities 
in the same direction, e.g., CU Cep ($-50$ km s$^{-1}$), IRC+60374 ($-52$ km s$^{-1}$), 
and AFGL 2999 ($-50$ km s$^{-1}$). 
Luminosity distances to these stars are inferred to be smaller than 3 kpc from their high IRAS flux densities.
In addition, MY Cep is an M supergiant with $V_{\rm lsr}=-56$ km s$^{-1}$ 
in the star cluster NGC 7419. The distance to this cluster has been well estimated to be
about 2.3 kpc from luminosities of member stars of the cluster (e.g., see \cite{bea94,sub06}). 
These example indicates that the stars  with $V_{\rm lsr}\sim -50$ km s$^{-1}$ do not necessarily 
belong to the Perseus arm, but they may be located much closely. 
Because the radial velocity expected by the galactic rotation is only $\sim -10$ km s$^{-1}$
at 1 kpc at $l=108^{\circ}$, and because the radial-velocity dispersion of stellar maser sources
is as small as $\sim 25$ km s$^{-1}$ at the solar neighborhood (see Appendix 2 of \cite{deg05}), 
IRAS 22480+6002 is possibly kinematically anomalous. 

%%%%%%%%%%%%%%%%%%%%%%%%%%%% subsection %%%%%%%%%%%%%%%%%%%%%%%
%%%%%%%%%%% Table 3 %%%%%%%%%%%%%%%%%%%
\begin{table*}[ht]
\begin{center}
\caption{Comparison of the catalogued positions of IRAS 22480$+$6002.} \label{tab:positions}
%\input{table3.tex}
%{\scriptsize
\footnotesize
\begin{tabular}{llllllll} \hline \hline    
Catalog  & Band & Assignment & epoch & R.A.(J2000) & decl..(J2000) & error & flux density \\ 
         &      &            &       &  \ h \ m \ s & \ \ ${\circ} \ \ ' \  \ ''$ & $''$  &  or magnitude     \\
 \hline    
   
  IRAS   & MIR & 22480+6002        & 1983   & 22 49 59.2   & +60 17 55    &  $11''\times 5''$ (19$^{\circ}$) &  $F_{\rm 12}=142$ Jy  \\
  MSX6   & MIR & G108.4255+00.8939 & 1995   & 22 49 58.89  & +60 17 56.8  &  0.3 & $F_{\rm C}=123$ Jy\\                         
         &     &                   &        &              &              &      &             \\
  2MASS  & NIR & 2495897+6017567   & 1997   & 22 49 58.97  & +60 17 56.8  &  0.29   & K=2.78      \\
  GSC1.2 & optical& 0426500695        & 1954 & 22 49 59.43  & +60 17 55.8  &  0.3 & R=12.29     \\
  GSC2.2 & optical& N012302336407     & 1989.6 & 22 49 58.900 & +60 17 57.17 &  0.3 & B=12.29     \\
         &     &                   &        &              &              &      &             \\
USNO-B1.0 & optical&   1502-0356025    & 1971.7 & 22 49 59.75  & +60 17 56.7  & (0.7, 1.0)     & R=8.87 \\
USNO-B1.0 & optical&   1502-0356023    & 1979.7 & 22 49 59.44  & +60 17 55.9  & (0.7, 0.2)     & R=8.73 \\
USNO-B1.0 & optical&   1502-0356019    & 1979.7 & 22 49 59.15  & +60 17 57.5  & (0.5, 0.7)     & B=12.63 \\
         &     &                   &        &              &              &      &             \\
JVN (this work) & radio &   22480+6002     & 2005.5 & 22 49 58.876 & +60 17 56.65 &  $0.1''$ &  $F_{\rm H_2O}\sim 8$ Jy  \\
 \hline
 
 \hline
 \end{tabular}
%}
 \end{center}
%\footnotemark[2] Mean full velocity width of a maser feature at half intensity. \\
\end{table*}

\subsection{Past optical/infrared data of IRAS 22480+6002 (=J22495897+6017568).}
Though this star is relatively bright at optical wavelengths ($V\sim 8.30$; Tyco Input catalog),
the star was not recorded in major optical catalogs, for example, not in Henry Draper (HD) Catalogue,
The Hipparcos and Tycho Catalogue, nor the General Catalog of Variable Stars, 
possibly because of  confusion by the nearby B5II star (TYC 4265-870-1; $V\sim 10.74$), 
located by about 12$''$ east. This was involved in The Washington Double Star 
Catalog\footnote{available at http://ad.usno.navy.mil/wds/wdstext.html.},
giving a separation of 10.9$''$ in 1901 and 12.0$''$ in 2006 with a small position angle variation
(by $\sim 7^{\circ}$) to the B star. From this data, we obtain the proper motion of 17 mas yr$^{-1}$ 
to the west for IRAS 22480+6002 relative to this B star. 
As noted by \citet{hum74}, this B5II star is probably not a binary counterpart
 because of the large radial velocity difference.
The ACT Reference Catalog gave a very small proper motion of this B5II star 
(less than 3 mas yr$^{-1}$); though The Hipparcos and Tycho Catalogue gave a large proper motion 
in declination due to position uncertainty, but this was corrected in ACT catalog.
We also checked the past catalogs recording the position of this star and summarized the results
in table 3.\footnote{
all the data except JVN are available in the VizieR database
({\it http://vizier.nao.ac.jp/viz-bin/VizieR}).} 
The GSC 1.2 catalog (which remeasured the POSS1 plate taken in 1950s) 
gave a different position by about 4.2$''$, which leads
a large proper motion of 84 mas yr$^{-1}$ if compared with GSC 2.0.
This value is much larger than the above-mentioned proper motion
computed from the Washington Double Star Catalog, though the proper motion vectors are roughly in the same direction. 
We believe the direct measurements of binary separation gives better values.
%(as far as the B5 counterpart does not show a large proper motion). 
Therefore, we adopt the proper motion of 17 mas yr$^{-1}$ for this star, and
get $U_0=-71$  km s$^{-1}$ and $V_0=-29$ km s$^{-1}$ for IRAS 22480+6002.
This motion is considerably peculiar for a population I disk star. 
%This spatial motion is close to the the Hercules stream group of K and M giants moves  
%with $<U_0>=-42.1\pm 26.8$ km s$^{-1}$ and $<V_0>=-50.6\pm 8.6$ km s$^{-1}$ 
%\citep{fam05}. Therefore, 
It is likely that IRAS 22480+6002 belongs to one of kinematical streaming groups
of stars as population II G and K giants \citep{fam05}. 
% Such peculiar motion of this group is considered to be made 
% by a dynamical effect due to the bulge bar \citep{fux01}, because the
% 0.73*3.1*10^21*17*0.4848*10^-8/3.16*10^7=59.0*10^6 cm/s=59 km/s
%  l=108  $U=170$ km s$^{-1}$ and $V=7 km s$^{-1}$
%ages of this group of stars are widespread \citep{fam05}. 

In the past, OH and H$_2$O  masers have been found 
in a few planetary and preplanetary nebulae \citep{zil89,mir01,sue07}, where
central stars of these objects have spectral types earlier than M. 
These masers are a remnant of circumstellar material which was ejected at the AGB phase of the central star.
The molecules responsible for masers are eventually to be dissociated.
In contrast, \citet{fix84} found OH 1665/1667 MHz emission toward several warm stars as RV Tau variables with spectral type of G and K,
but so far only one case (TW Aql, a semi-regular variable of K7III) was confirmed 
to be a circumstellar maser \citep{pla91}. 
SiO masers, which are emitted within a few stellar radii of the central star
(much closer than H$_2$O masers are emitted), 
were not detected in these warm objects before. 
An exceptional case is the RV Tau variable, R Sct, with spectral type K0Ib. 
This object exhibits strong SiO and weak H$_2$O masers (I. Yamamura, 2004 private communication) 
as well as the 4 $\mu$m SiO first overtone bands \citep{mat02}. 
The RV Tau variables are believed to be low-mass post-AGB stars \citep{jul86} with
low metal abundances (population II; \cite{gir00}), though these are spectroscopically 
classified as supergiants. Their spectral types change between K and M-type with light variation \citep{pol97}. 
The atmosphere of late-type supergiants are not 
in hydrostatic equilibrium; effective temperature increases with decreasing metalicity \citep{lev05}. 
The RV Tau variables are enshrouded by dust shell, and  
CO emission has been detected in two of these variables \citep{buj88}.
Although  the optical counterpart of IRAS 22480+6002 is not reported to show any strong light variability 
(e.g., TASS; The Amateur Sky Survey\footnote{data available at http://www.tass-survey.org/}), 
the optical spectroscopic classification, middle infrared properties,  and maser characteristics
of IRAS 22480+6002 indicate a close similarity to the properties of the RV Tau variables. 

%%%%%%%%%%%%%%%%%%%%%%%%%%%%%%%% summary %%%%%%%%%%%%%%%%%%%%%%
\section{Summary}  
We observed the H$_2$O maser emission distribution around the evolved star IRAS 22480+6002 
with the Japanese VLBI Network. The maser emission was found to come 
from an area elongated in the east-west direction by about 50 mas, 
but no clear trend was found between radial velocities and spatial positions.
With the three-epoch two-month interval VLBI observations, we found relatively large proper motions
of a few up to 5 mas yr$^{-1}$ for individual H$_2$O maser features. They exhibits no systematic trend 
in their velocity field, and the inner envelope of this star is very disturbed.  
Applying the statistical parallax method, we obtained a distance of $\sim 1.0 \pm 0.4 $ kpc,
which is significantly smaller than the value previously thought.
This distance gives a small luminosity of $\sim 6 \times 10^3 \ L_{\odot}$, but not unreasonably small 
as a K5 or M0-type supergiant. Combination of the distance 
and the previous optical proper motion data of this star place this
object to be in a dynamically streaming group of stars. The maser characteristic, and optical and infrared 
properties of this star are quite similar to those of population II post-AGB stars as RV Tau variables.     

\bigskip
We acknowledge all staff members and students who have helped in array 
operation and in data correlation of the JVN/VERA.  H.~I. was supported by 
Grant-in-Aid for Scientific Research from Japan Society for Promotion 
Science (18740109). This research was made use of VizieR database, operated at CDS, 
Strasbourg, France.

\

%%%%%% appendix error estimation
% \include{error_est}
% note  1 mas/yr at 1 kpc= 3.086*10^21 * 0.4848 10^-8 /(3.156 10^7) =4.740 km/s 

%\clearpage

% \clearpage
% \twocolumn

%%%%%%%%%%%%%%%%%%%%%%%%%%%%%%%%%%%%%%%%%
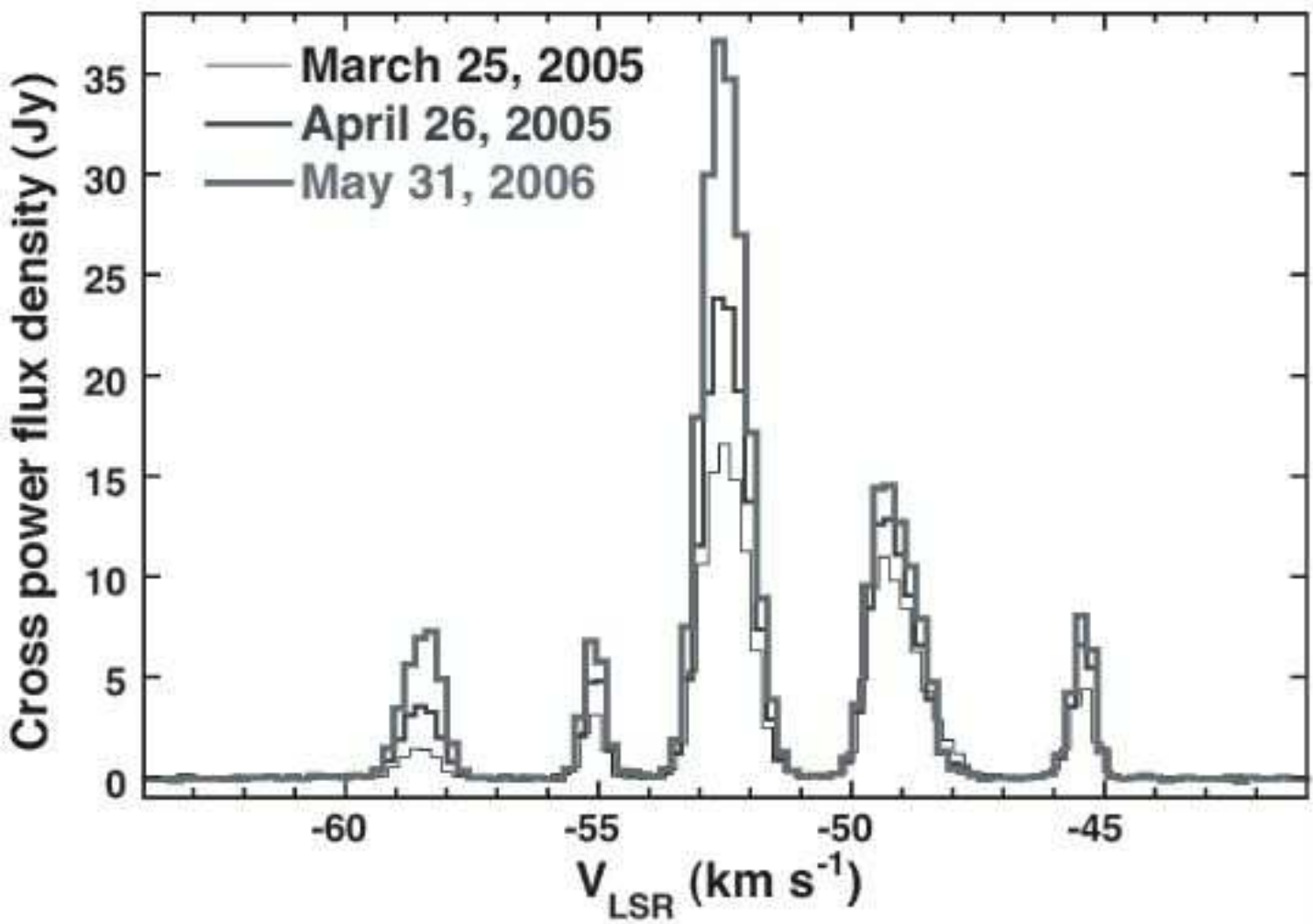
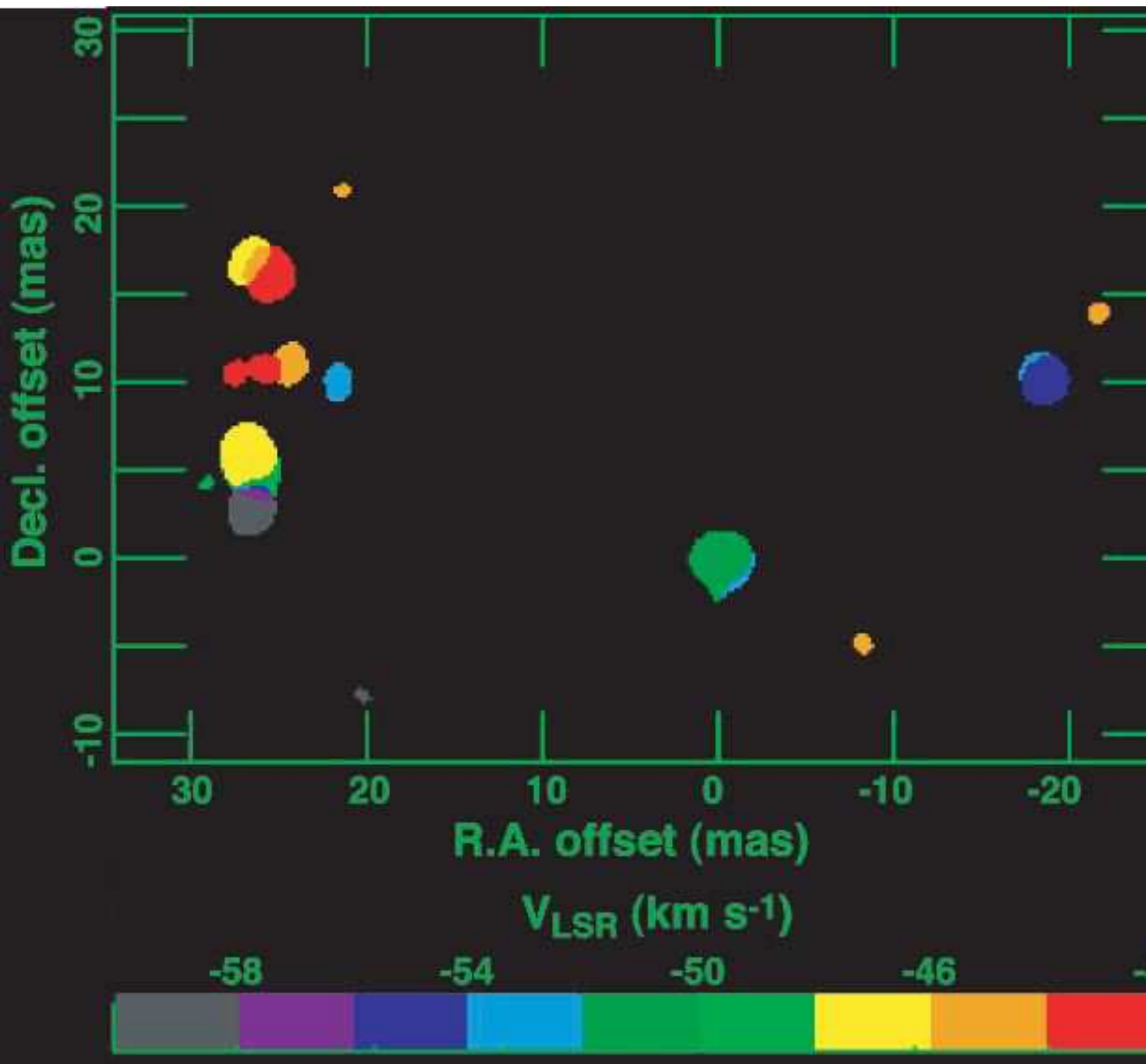
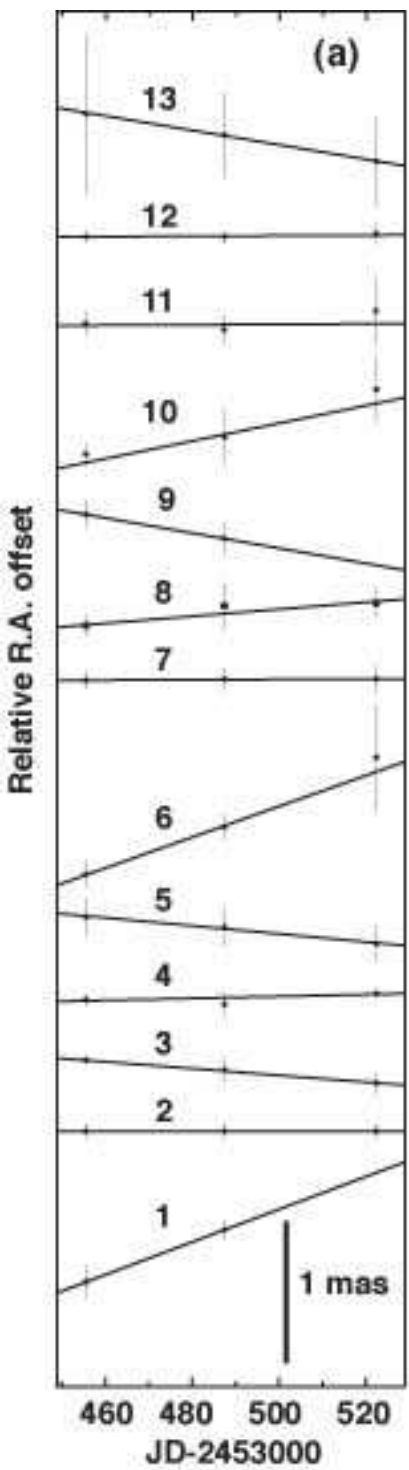
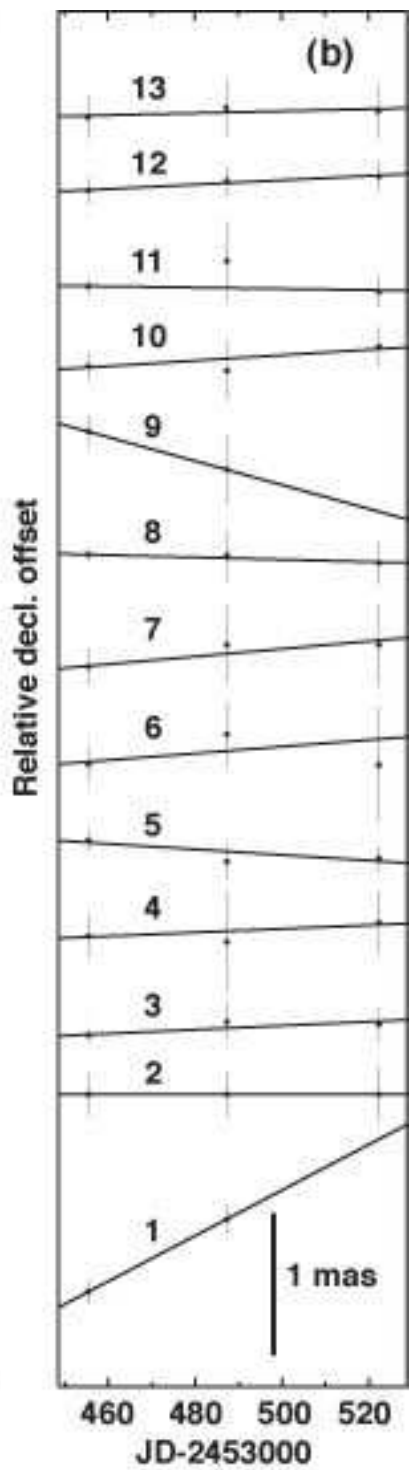
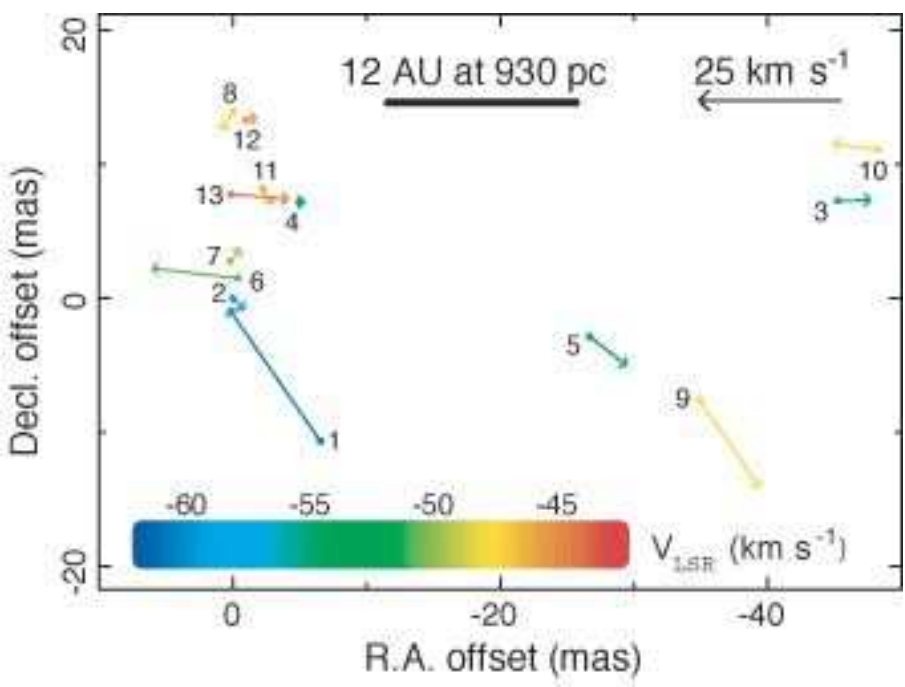
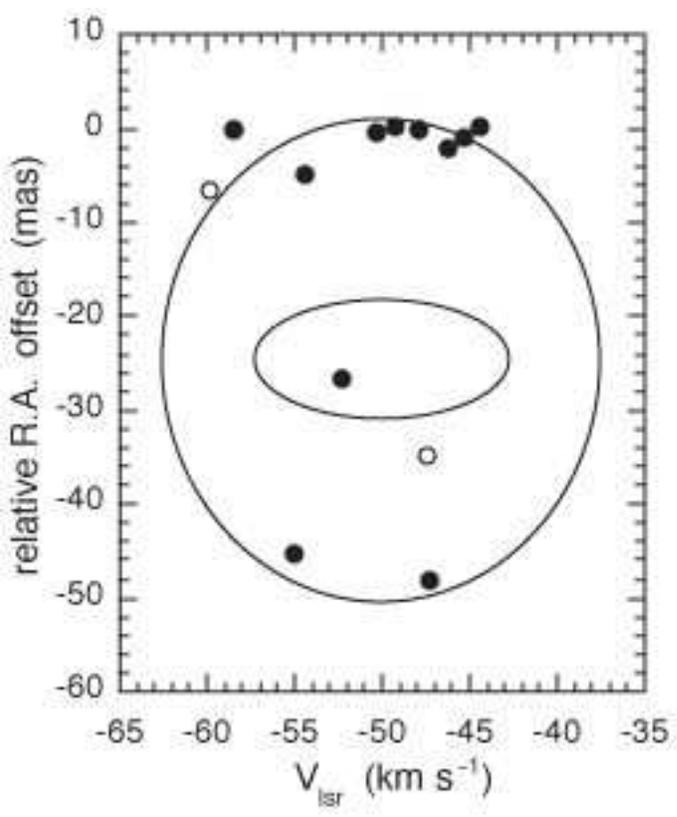
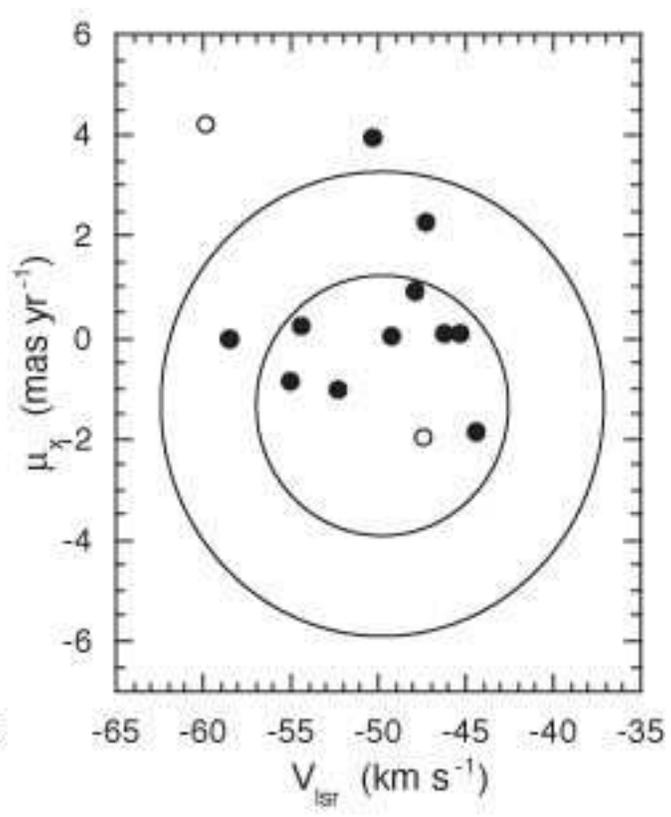
\end{document}